\documentclass{article}

     \PassOptionsToPackage{numbers, compress}{natbib}

\usepackage[final]{neurips_2021_ml4ps}
\usepackage{tabularx,colortbl}
\usepackage[numbers]{natbib}
\usepackage{graphicx}
\usepackage{amssymb}
\usepackage{pifont}
\usepackage{amsmath}
\usepackage{wrapfig,booktabs}

\usepackage[utf8]{inputenc} 
\usepackage[T1]{fontenc}    
\usepackage{hyperref}       
\usepackage{url}            
\usepackage{booktabs}       
\usepackage{amsfonts}       
\usepackage{nicefrac}       
\usepackage{microtype}      
\usepackage{xcolor}         

\usepackage{todonotes}
\definecolor{orange}{rgb}{1,0.5,0}
\definecolor{red}{rgb}{1, 0, 0}

\newcommand{\cmark}{\ding{51}}%
\newcommand{\xmark}{\ding{55}}%
\title{Deep-SWIM: A few-shot learning approach to classify Solar WInd Magnetic field structures}
\author{
  Hala Lamdouar \\
  University of Oxford \\
  \texttt{lamdouar@robots.ox.ac.uk}
  \And
  Sairam Sundaresan \\
  Intel Labs\\
  \texttt{sairam.sundaresan@intel.com} \\
  \And
  Anna Jungbluth \\
  University of Oxford \\
  \texttt{anna.jungbluth@physics.ox.ac.uk} \\
  \And
  Sudeshna Boro Saikia \\
  University of Vienna \\
  \texttt{sudeshna.boro.saikia@univie.ac.at} \\
  \And
  Amanda Joy Camarata \\
  Colorado School of Mines \\
  \texttt{ajcamarata@mines.edu} \\
  \And
  Nathan Miles \\
  University of California,Los Angeles \\
  \texttt{ndmiles@g.ucla.edu} \\
  \And 
  Marcella Scoczynski \\
  Federal University of Technology – Paran\'a \\
  \texttt{marcella@utfpr.edu.br} \\
  \And
  Mavis Stone \\
  Harvey Mudd College \\
  \texttt{mstone@g.hmc.edu} \\
  \And
  Anthony Sarah \\
  Intel Labs\\
  \texttt{anthony.sarah@intel.com} \\
  \And
  Andr\'es Mu\~noz-Jaramillo \\
  Southwest Research Institute\\
  \texttt{amunozj@boulder.swri.edu} \\
  \And
  Ayris Narock \\
  NASA Goddard Space Flight Center \\
  ADNET Systems Inc \\
  \texttt{ayris.a.narock@nasa.gov} \\
  \And
  Adam Szabo \\
  NASA Goddard Space Flight Center \\
  \texttt{adam.szabo-1@nasa.gov} \\
}

\begin{document}

\maketitle

\begin{abstract}


The solar wind consists of charged particles ejected from the Sun into interplanetary space and towards Earth. Understanding the magnetic field of the solar wind is crucial for predicting future space weather and planetary atmospheric loss. 
Compared to large-scale magnetic events, smaller-scale structures like magnetic discontinuities are hard to detect but entail important information on the evolution of the solar wind.
A lack of labeled data makes an automated detection of these discontinuities challenging.
We propose Deep-SWIM, an approach leveraging advances in contrastive learning, pseudo-labeling and online hard example mining to robustly identify discontinuities in solar wind magnetic field data. Through a systematic ablation study, we show that we can accurately classify discontinuities despite learning from only limited labeled data.
Additionally, we show that our approach generalizes well and produces results that agree with expert hand-labeling.



\end{abstract}


\section{Introduction}
\label{introduction}


Every star has an extended atmosphere of charged particles reaching far into interplanetary space. For the Sun, this extended atmosphere is called the solar wind and it significantly impacts Earth's space weather. Understanding the magnetic field that drives the solar wind and the different magnetic structures it contains is essential to predicting the effect of the solar wind on Earth and future extraterrestrial missions. Furthermore, knowledge on solar wind properties helps constrain our host star’s impact on planetary atmospheres to inform our understanding of exoplanet habitability.
Over the past decades, instruments like the WIND spacecraft have collected measurements of the magnetic field of the solar wind. While some magnetic events, e.g. interplanetary coronal mass ejections, are high-strength and therefore easy to detect \cite{dosSantos2020,nguyen2019}, smaller-scale events, like \textit{\textbf{discontinuities}}, are more difficult to spot. Discontinuities describe "discontinuous" spatial changes in the solar magnetic field. 
Although they only last for a few seconds or a few minutes, and are usually weak in amplitude, the occurence rate of discontinuties is quite high, ranging up to 10,000s over a few months \citep{malaspina2012}. 
Experts who can identify these magnetic structures are limited, and hand-labeling is too time-consuming to catalog all available data.

Applying machine learning (ML) to solar wind magnetic field data allows us to quickly and accurately classify these small-scale structures to create a comprehensive catalog of all measured discontinuities. To do this, we employ both supervised and semi-supervised machine learning approaches using 1D and 2D convolutional neural networks (CNNs). As the availability of labeled data is limited, and the classes are heavily imbalanced, we apply various data augmentation techniques and leverage recent advances in contrastive learning~\cite{khosla2020supervised}, pseudo-labeling~\cite{Lee2013pseudolabeling}, and online hard example mining~\cite{shrivastava2016training}. We show that contrastive learning and pseudo-labeling significantly improve the performance of the model and allow it to generalize better to unseen data.

\section{Data sets}
\label{datasets}

We use data from the Magnetic Field Investigation (MFI) fluxgate magnetometer on board the WIND spacecraft \cite{Lepping1995,Koval2013} \footnote{The data is publicly available on NASA's Coordinated Data Analysis Web at \href{https://cdaweb.gsfc.nasa.gov/}{https://cdaweb.gsfc.nasa.gov/}}. 
WIND is positioned in a halo orbit around Lagrange point 1, measuring small-scale magnetic structures embedded in the solar wind \cite{szabo2019long} at a constant distance from the Sun and at sub-second resolution.
To remove artifacts of the 3s rotational period of the instrument, we smooth and re-sample the original 11 Hz data to a 3s cadence. 
The time series are split into 5-minute intervals, and the x, y, and z components of the magnetic field are stacked as inputs to the CNN. 
For our supervised approach we assign binary labels to each segment specifying whether it contains a discontinuity or not. 
The labels are provided by expert hand-labeling for one day (11-18-2018) or using a non-machine learning based heuristic approach which compares the rotation angle of neighboring points to identify discontinuities (2006-2021) \footnote{The hand-labeled and heuristic catalog were provided by Dr. Adam Szabo and Dr. Ayris Narock who are both authors of this work.}. 
Importantly, not all labels in the heuristic catalog are expected to be correct. For the one day of joint labeling, both the heuristic approach and hand-labeling identified 39 discontinuities out of which 24 discontinuities were identified by both approaches.
In addition to noise in the labeling, our data set is highly imbalanced. Only around 15\% of our 5-minute intervals contain discontinuities.
To perform our ablation study, we select a representative sample of 3 months of data (May - July 2018). We split the data using stratified sampling to ensure the correct proportional representation of positive and negative examples and randomly allocate segments to the training and validation sets. 
20\% of the data (5428 5s segments) was used for training the supervised component of our approach, 70\% (18492 5s segments) were used for training the semi-supervised component, and 10\% (2576 5s segments) was used for validation. The one day of expert hand-labeled data (11-18-2018, 286 5s segments) was used as a final test set.

\section{Methodology} 
\label{method}

\begin{figure}[t!]
    \small
    \begin{center}
    \includegraphics[width=1\textwidth]{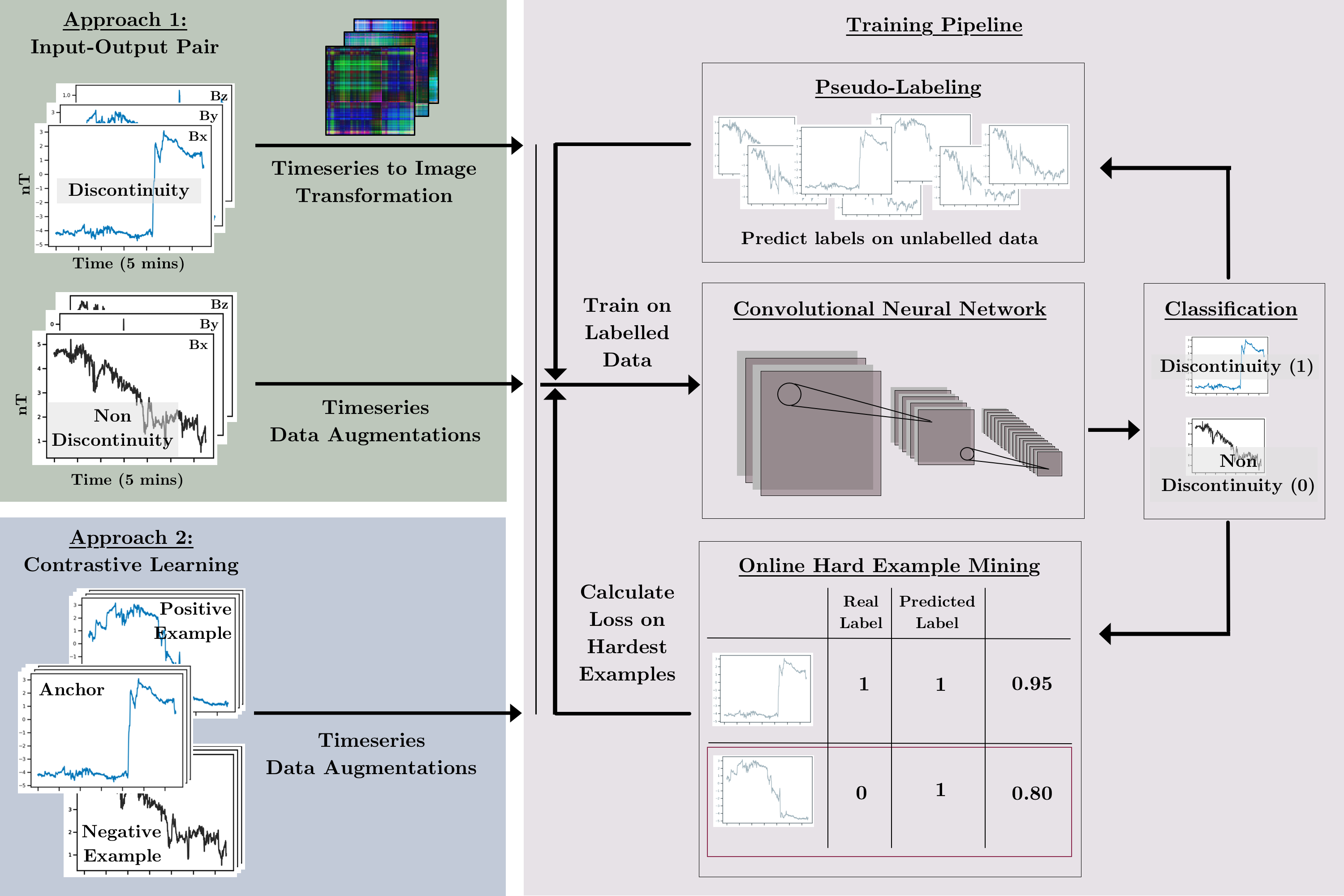}
    \caption{\textbf{Schematic of our machine learning pipeline}. We compare and evaluate two modalities (1) a supervised approach using a standard data-label pairing as input, and (2) a semi-supervised approach using contrastive learning where the input data consists of an anchor (data of interest), a positive example, and a negative example. Additional model improvements are achieved by employing pseudo-labeling and online hard example mining during training.}
    \label{fig:pipeline}
    \end{center}
\end{figure}

A schematic of our ML pipeline is shown in Figure \ref{fig:pipeline}. We compare the performance of our solar wind magnetic field classification based on two modalities.
Firstly, we use a standard data-label pairing as input to a four layer 1D-CNN. We also test the performance of 2D-CNN with a ResNet-18 backbone after converting the original time series into images using a Gramian Angular Field (GAF) conversion \cite{wang2015imaging}. The successful use of GAF conversions for ml-based astrophysics research has been demonstrated e.g. in \cite{johnson2020rotnet}.
While we evaluated other backbone architectures including AlexNet, Inception and larger ResNets, we found that they either quickly overfit the data or do not converge.
Secondly, we employ contrastive learning \cite{10.1109/CVPR.2005.202} using three examples as simultaneous inputs to the model. In this case, the classification head of each model is replaced with a fully connected layer which has a size equal to the size of the embedding dimension desired. A downstream classifier head is appended to this for fine tuning to produce classification results once the contrastive model has been properly trained.
The training pipeline is made increasingly more sophisticated through the additions of online hard example mining and pseudo-labeling.

\textbf{Data Augmentations. }
To mitigate limitations of our imbalanced data set, we employ data augmentations to increase the number of examples containing discontinuities. We perform horizontal mirroring, random scaling between $0.5$ and $2$, random inverting, and random channel shuffling of the time series data. These data augmentations all preserve the physics of the solar magnetic field and present scenarios that could realistically be measured by the instrument; for instance, an inverted magnetic field could be observed during a different solar cycle with reversed magnetic poles.

\textbf{Contrastive Learning. }
By employing contrastive learning \cite{10.1109/CVPR.2005.202,khosla2020supervised}, rather than learning from a single input-output pair, the model aims to learn the similarities and differences between examples. For this, three examples are paired; the anchor ($x$), i.e. the example of interest, a positive example of the same class as the anchor ($x^+$), and a negative example from a different class ($x^-$). The model learns an embedding space to group similar examples close together and dissimilar examples far apart. To do this, we employ a triple margin loss \cite{Schroff_2015,BMVC2016_119}.

\begin{equation}
    L_{triple}(x, x^+, x^-) = \sum \max(0, \|f(x) - f(x^+)\|^2 - \|f(x) - f(x^-)\|^2 + \alpha),
\end{equation}

where $\alpha$ is the margin between the positive and negative pairs (here, $\alpha=1$), and $f$ is the learned embedding function. Further for the triplet margin loss, the positive and negative examples are chosen at random.

\textbf{Pseudo-Labeling. }
Pseudo-labeling \cite{Lee2013pseudolabeling} is a semi-supervised learning approach to maximize the potential of small labeled data sets. The model is first trained on limited labeled data and subsequently used to predict labels of unlabeled data. The pseudo labels are refreshed every epoch. The total loss is calculated as the sum of the loss of the labeled and unlabeled data ($\mathrm{L_{Total} = L_{labeled} + \alpha \cdot L_{unlabeled}}$) with a weighting factor $\alpha$ that changes with epoch $t$ according to equation \ref{eq:alpha}. Here, $\alpha_f = 3$, $T_1 = 5$, and $T_2 = 100$.

\begin{equation} \label{eq:alpha}
    \alpha(t) =  
\begin{cases}
    0, & t < T_1 \\
    \frac{t-T_1}{T2-T1}\alpha_f, & T_1 < t < T_2 \\
    \alpha_f, & t > T_2
\end{cases}
\end{equation}

\textbf{Online Hard example mining. }
Most machine learning tasks have combinations of easy and hard examples to learn from. Online Hard example mining (OHEM) \cite{shrivastava2016training} focuses on the model's performance on examples that are either misclassified
with high confidence or correctly classified with reduced confidence. The total loss is calculated as $\mathrm{L_{total} = \omega \cdot L_{OHEM} + (1 - \omega) \cdot L_{raw}}$, where $\mathrm{L_{raw}}$ is the standard loss across all samples, $\mathrm{L_{OHEM}}$ is the loss computed on the top $70\%$ of hardest examples, i.e.\ examples with the highest loss values, and $\omega$ is a weighting factor. In our experiments, $\omega$ was set to $0.8$.

\section{Experiments}
\label{experiments}

\begin{wrapfigure}[37]{R}{0.4\textwidth}
\centering
\includegraphics[width=0.4\textwidth]{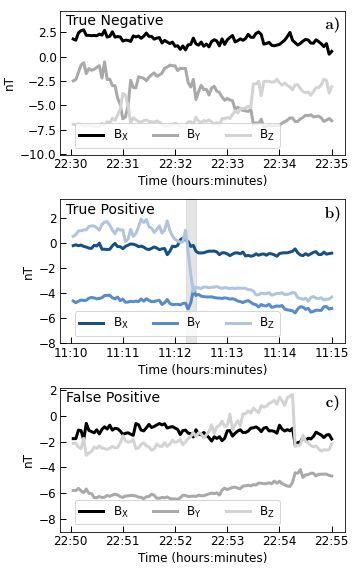}
\caption{\textbf{Example performance of our best model.} The model was tested on one day of expert hand-labeled data (11-18-2018). Here we show a) a true negative, b) true positive, and c) false positive example of the model's labeling. The colors in each graph present the x, y, and z component of the solar wind magnetic field. The gray shading in panel b) shows the time stamp of the discontinuity.}
\label{fig:examples}
\end{wrapfigure}
\paragraph{Training. }
In this section, we compare our experiments to a non deep learning baseline using Support-Vector Machines (SVMs)~\cite{boser1992training,cortes1995support}.
The SVM
baseline is obtained using a linear kernel with $\gamma = 1.0$, $c = 0.1$. These parameters are chosen through a grid search over the parameter space and selecting the combination which yields the best Area Under the receiver operating characteristic Curve (AUC) score on the validation set. We have also tested SVM with a Radial Basis Function (RBF) kernel as well as polynomial and linear and found the linear kernel to provide the best performance. Note that we apply the SVM directly to time series data, but extract Oriented Rotated BRIEF (ORB) features~\cite{rublee2011orb} from images prior to using the SVM.
Both our time-series and image based CNNs are trained on a single NVIDIA A100 GPU and use the Adam optimizer, a learning rate of $10^{-3}$, and a batch size of 16. These parameters are obtained through hyper-parameter search runs. Additionally, the 2D-CNN is initialized with ImageNet weights to facilitate transfer learning.

\paragraph{Results and Discussion.}
We conduct an ablation study on the different components of our training pipeline and present the results in Table~\ref{tab:ablation}. We find that the SVM approach, c.f.\ experiments (A1) and (B1), provides considerably lower performance compared to CNN-based approaches, especially for images combined with ORB features that fail to extract distinguishing attributes from the Gramian Angular Field conversions.
The raw time series and image data types yield comparable results when the CNNs without additions are considered, with slightly better performance for the raw time series. However, once we introduce pseudo-labeling, raw times series (B4) clearly outperform image data types (A4) by over $27 \%$ AUC. Moreover, we found that using contrastive learning did not alleviate the convergence issues in the image based models while pseudolabeling resulted in divergence. Overall, we achieve the best performance, i.e.\ experiment (B5) with $90\%$ AUC, using a combination of contrastive learning, pseudo-labeling, and OHEM. While the AUC scores for B3 to B7 are close, we found qualitatively, B3 appeared to be robust on certain hard examples. In our future work, we plan to substantiate this with a statistical significance test.
All experiments presented in Table~\ref{tab:ablation} are run with a random seed of $42$. Varying the random seed of our best model across 2 orders of magnitude resulted in the following median absolute deviations (MAD): $\mathrm{{MAD}_{Precision}}=0.014$, $\mathrm{{MAD}_{Recall}}=0.000$ and $\mathrm{{MAD}_{AUC}}=0.003$.

To further evaluate our best model, we compare the model's labels to the expert hand-labeled data set (11-18-2018). An example of a true negative, true positive, and false positive labeling is provided in Figure~\ref{fig:examples}. The false positive labeling is caused by the sharp dip in the $\mathrm{B_Z}$ component around $10:54$ pm, which is incorrectly understood to be a discontinuity. Our best model achieves a precision of $0.52$, a recall of $0.73$, and an AUC of $0.82$ on this final test set. 


\begin{table}[!htb]
\footnotesize
\centering
\setlength{\tabcolsep}{3.5pt}
\begin{tabular}{c|c|c|ccc|ccc}
\cline{2-9}  & Architecture & data type & Contrastive Learning & Pseudo-labeling & OHEM & Precision  & Recall & AUC\\
\hline A1 & SVM & images & \xmark &  \xmark & \xmark & 0.00 & 0.00  & 0.50  \\
A2 & ResNet-18 & images & \xmark &  \xmark & \xmark & 0.51 & 0.65  & \bf{0.79}  \\
A3 & ResNet-18 & images & \xmark &  \cmark & \xmark & 1.00 & 0.03  & 0.51  \\
A4 & ResNet-18 & images & \xmark &  \cmark & \cmark & 0.8 & 0.26  & 0.62  \\

\midrule
B1 & SVM & time series & \xmark &  \xmark & \xmark & 1.00 & 0.23  & 0.62  \\
B2 & 1D-CNN & time series & \xmark &  \xmark & \xmark & 0.66 & 0.61  & 0.79 \\
B3 & 1D-CNN & time series & \xmark &  \cmark & \xmark & 0.63 & 0.84  & 0.89  \\
B4 & 1D-CNN & time series & \xmark &  \cmark & \cmark & 0.67 & 0.84  & 0.89  \\
B5 & 1D-CNN & time series & \cmark &  \cmark & \cmark & 0.59 & 0.87  & \bf{0.9}   \\
B6 & 1D-CNN & time series & \cmark &  \cmark & \xmark & 0.47 & 0.90  & 0.89  \\
B7 & 1D-CNN & time series & \cmark &  \xmark & \xmark & 0.77 & 0.77  & 0.87   \\

\bottomrule
\end{tabular}
\caption{Ablation study of our proposed approach to classify discontinuities. We select the best model based on the higher AUC (Area Under the receiver operating characteristic Curve).}
\label{tab:ablation}
\end{table}
\vspace{2cm}
\section{Conclusions}
\label{conclusions}
In this work, we present a few-shot learning approach for classifying the magnetic structures of solar wind magnetic field data. Specifically, we focus on identifying discontinuities, which present "discontinuous" spatial changes in the magnetic field. 
Our proposed approach combines contrastive learning and pseudo-labeling to overcome the large imbalance of our data set. 
We show that incorporating online hard example mining further improves the overall performance. Finally, our method, trained on heuristic labels, generalizes well to expert hand-labeling, circumventing the labor-intensive and time-consuming process of manual annotation of such complex data.


\section*{Broader Impact}
\label{impact}

At the present moment, state-of-the-art applications of deep learning to solar wind measurements are primarily confined to supervised classification of large-scale structures like interplanetary coronal mass ejections (ICME) \cite{nguyen2019,dosSantos2020}. 
The primary driver behind this is the occurrence rate of these structure. Large-scale structures like ICMEs occur at a rate of about 10 to 50 times per year depending on solar activity and are very pronounced and easily detectable in in-situ measurements. 
While this facilitates the creation of hand-labeled data sets, there are still issues associated with misclassification of these structures due to human bias \cite{russell_2005_icme}. 
For small-scale structures like  discontinuities, creating a hand-labeled catalog is infeasible since discontinuities occur at rates of more than 10,000 times per year. 
Here we show that we can apply pseudo-labeling to leverage the potential of small data sets and accurately identify these target structures.
For future research, this enables to hand-label small amounts of data, or apply costly heuristic approaches to only a subset of the data to create a small training set for the supervised component of our proposed approach. 



\onecolumn
\section*{Acknowledgments and Disclosure of Funding}
\label{acknowledgments}
This work was conducted at the Frontier Development Laboratory (FDL) USA 2021. The FDL USA is a public / private research partnership between NASA, the SETI Institute and private sector partners including Google Cloud, Intel, IBM, Lockheed Martin, and NVIDIA. These partners provide the data, expertise, training, and compute resources necessary for rapid experimentation and iteration in data-intensive areas.


\bibliography{main_bib}

\begin{thebibliography}{10}

\bibitem{dosSantos2020}
Luiz~FG dos Santos, Ayris Narock, Teresa Nieves-Chinchilla, Marlon Nu{\~n}ez,
  and Michael Kirk.
\newblock Identifying flux rope signatures using a deep neural network.
\newblock {\em Solar Physics}, 2020.

\bibitem{nguyen2019}
Gautier Nguyen, Nicolas Aunai, Dominique Fontaine, Erwan~Le Pennec, Joris~Van
  den Bossche, Alexis Jeandet, Brice Bakkali, Louis Vignoli, and Bruno
  Regaldo-Saint Blancard.
\newblock Automatic detection of interplanetary coronal mass ejections from in
  situ data: A deep learning approach.
\newblock {\em The Astrophysical Journal}, 2019.

\bibitem{malaspina2012}
David~M. {Malaspina} and J.~T. {Gosling}.
\newblock {Two spacecraft observations of magnetic discontinuities in the solar
  wind with STEREO}.
\newblock {\em Journal of Geophysical Research (Space Physics)},
  117(A4):A04109, April 2012.

\bibitem{khosla2020supervised}
Prannay Khosla, Piotr Teterwak, Chen Wang, Aaron Sarna, Yonglong Tian, Phillip
  Isola, Aaron Maschinot, Ce~Liu, and Dilip Krishnan.
\newblock Supervised contrastive learning.
\newblock In {\em Proceedings of the Conference on NeuralInformation Processing
  Systems (NeurIPS)}, 2020.

\bibitem{Lee2013pseudolabeling}
Dong-Hyun Lee.
\newblock Pseudo-label : The simple and efficient semi-supervised learning
  method for deep neural networks.
\newblock In {\em Proceedings of the International Conference on Machine
  Learning (ICML) Workshop : Challenges in Representation Learning (WREPL)},
  2013.

\bibitem{shrivastava2016training}
Abhinav Shrivastava, Abhinav Gupta, and Ross Girshick.
\newblock Training region-based object detectors with online hard example
  mining.
\newblock In {\em Proceedings of the Conference on Computer Vision and Pattern
  Recognition (CVPR)}, 2016.

\bibitem{Lepping1995}
R.~P. Lepping, M.~H. Ac{\~{u}}na, L.~F. Burlaga, W.~M. Farrell, J.~A. Slavin,
  K.~H. Schatten, F.~Mariani, N.~F. Ness, F.~M. Neubauer, Y.~C. Whang, J.~B.
  Byrnes, R.~S. Kennon, P.~V. Panetta, J.~Scheifele, and E.~M. Worley.
\newblock The {WIND} magnetic field investigation.
\newblock {\em Space Science Reviews}, 1995.

\bibitem{Koval2013}
Andriy Koval and Adam Szabo.
\newblock Magnetic field turbulence spectra observed by the wind spacecraft.
\newblock {\em Proceedings of the American Institute of Physics Conference
  (AIP)}, 2013.

\bibitem{szabo2019long}
Adam Szabo, Andriy Koval, and Ayris Narock.
\newblock Long-term observations of interplanetary discontinuities by the wind
  spacecraft.
\newblock In {\em American Geophysical Union (AGU) Fall Meeting Abstracts},
  2019.

\bibitem{wang2015imaging}
Zhiguang Wang and Tim Oates.
\newblock Imaging time-series to improve classification and imputation.
\newblock In {\em Proceedings of the International Joint Conference on
  Artificial Intelligence (IJCAI)}, 2015.

\bibitem{johnson2020rotnet}
J.~Emmanuel Johnson, Sairam Sundaresan, Tansu Daylan, Lisseth Gavilan,
  Daniel~K. Giles, Stela~Ishitani Silva, Anna Jungbluth, Brett Morris, and
  Andrés Muñoz-Jaramillo.
\newblock Rotnet: Fast and scalable estimation of stellar rotation periods
  using convolutional neural networks.
\newblock In {\em Proceedings of the Conference on Neural Information
  Processing Systems (NeurIPS) Workshop on Machine Learning and the Physical
  Sciences}, 2020.

\bibitem{10.1109/CVPR.2005.202}
Sumit Chopra, Raia Hadsell, and Yann Lecun.
\newblock Learning a similarity metric discriminatively, with application to
  face verification.
\newblock In {\em Proceedings of the Conference on Computer Vision and Pattern
  Recognition (CVPR)}, 2005.

\bibitem{Schroff_2015}
Florian Schroff, Dmitry Kalenichenko, and James Philbin.
\newblock Facenet: A unified embedding for face recognition and clustering.
\newblock In {\em Proceedings of the Conference on Computer Vision and Pattern
  Recognition (CVPR)}, 2015.

\bibitem{BMVC2016_119}
Daniel~Ponsa Vassileios~Balntas, Edgar~Riba and Krystian Mikolajczyk.
\newblock Learning local feature descriptors with triplets and shallow
  convolutional neural networks.
\newblock In {\em Proceedings of the British Machine Vision Conference (BMVC)},
  2016.

\bibitem{boser1992training}
Bernhard~E Boser, Isabelle~M Guyon, and Vladimir~N Vapnik.
\newblock A training algorithm for optimal margin classifiers.
\newblock In {\em Proceedings of the Annual Workshop on Computational Learning
  Theory (AWCLT)}, 1992.

\bibitem{cortes1995support}
Corinna Cortes and Vladimir Vapnik.
\newblock Support-vector networks.
\newblock {\em Machine learning}, 1995.

\bibitem{rublee2011orb}
Ethan Rublee, Vincent Rabaud, Kurt Konolige, and Gary Bradski.
\newblock \uppercase{ORB}: An efficient alternative to sift or surf.
\newblock In {\em Proceedings of the International Conference on Computer
  Vision (ICCV)}, 2011.

\bibitem{russell_2005_icme}
C.~T. {Russell} and A.~A. {Shinde}.
\newblock {On Defining Interplanetary Coronal Mass EJECTIONs from Fluid
  Parameters}.
\newblock {\em Solar Physics}, 2005.

\end{thebibliography}

\end{document}